\documentclass[12pt]{article}
\usepackage[backend=biber,style=apa]{biblatex}
\usepackage{tabularx}
\usepackage{booktabs}
\usepackage{graphicx}
\usepackage{array}
\usepackage{lipsum} 
\usepackage{graphicx}
\usepackage{float}
\usepackage[a4paper, margin=1in]{geometry}
\usepackage{setspace}
\usepackage{hyperref}
\usepackage{authblk}
\begin{document}

\title{From Bias to Accountability: How the EU AI Act Confronts Challenges in European GeoAI Auditing}

\author{Natalia Matuszczyk}
\author{Craig R. Barnes}
\author{S. Roy}
\author{Rohit Gupta}
\author{Bulent Ozel}
\author{Aniket Mitra}
\affil{Computationele Wetenschapsgroep (CW) \\
hello@computationelewetenschapsgroep.com}

\maketitle
\begin{abstract}
Bias in geospatial artificial intelligence (GeoAI) models has been documented, yet the evidence is scattered across narrowly focused studies. We synthesize this fragmented literature to provide a concise overview of bias in GeoAI and examine how the EU’s Artificial Intelligence Act (EU AI Act) shapes audit obligations. We discuss recurring bias mechanisms, including representation, algorithmic and aggregation bias, and map them to specific provisions of the EU AI Act. By applying the Act's high-risk criteria, we demonstrate that widely deployed GeoAI applications qualify as high-risk systems. We then present examples of recent audits along with an outline of practical methods for detecting bias. As far as we know, this study represents the first integration of GeoAI bias evidence into the EU AI Act context, by identifying high-risk GeoAI systems and mapping bias mechanisms to the Act's Articles. Although the analysis is exploratory, it suggests that even well-curated European datasets should employ routine bias audits before 2027, when the AI Act’s high-risk provisions take full effect. \end{abstract}

\newpage

\section{Introduction}

Geospatial artificial intelligence (GeoAI) now influences decisions about flood defense placement, measuring climate change impacts, and public health policy \autocite{mahmood2022strategic}. Recent audits show rural buildings missed by global footprint datasets \autocite{gevaert2024auditing}, and crime-prediction tools that perpetuate discrimination \autocite{richardson2019dirty}. Despite this, European spatial data remain largely unexamined. Recent isolated audits of global land cover maps \autocite{venter2024uncertainty} and Copernicus imperviousness data \autocite{strand2022impervioussness} confirm representation bias, reveal significant estimation biases and geographic inaccuracies, yet each focuses on a single theme and stops short of engaging with regulatory implications. 

Since August 2024, the EU Artificial Intelligence Act (AI Act) has imposed tiered obligations on AI providers. Most GeoAI applications related to critical infrastructure, public services or law enforcement fall into the high-risk tier of Annex III. Articles 9-15 require that high-risk AI systems use representative data, document known and residual error, and ensure human oversight. Non-compliance can result in fines of up to 7\% of global turnover. Yet publicly documented audits are still scarce, which this paper examines. 

Despite this regulatory urgency, the literature reveals substantial gaps. Most studies on geospatial bias are either domain‑specific (e.g. building‑footprint accuracy) or technology‑specific (e.g. geoparsing), leaving no accessible overview of how and why bias arises in GeoAI. While discrete audits now exist for individual INSPIRE and Copernicus layers, our study is the first to integrate those findings and link them to provisions of the EU AI Act. Specifically, this paper addresses existing gaps by:

\begin{itemize}
    \item Providing an overview of evidence for bias in geospatial AI, including implications for European datasets
    \item Demonstrating how Articles 9-15 of the EU AI Act apply to the common types of bias in GeoAI
    \item Offering practical examples of GeoAI audits to guide practitioners in identifying and mitigating bias.
\end{itemize}

Together, these goals tackle the core puzzle: why, when bias is widely acknowledged, do routine GeoAI audits remain the exception rather than the norm? To achieve these objectives, the paper is structured as follows: Section 2 provides background on GeoAI and bias in machine learning. Section 3 discusses the evidence for the research gap in European data. Section 4 outlines the legal framework of the EU AI Act. Section 5 briefly touches upon the future outlook of General Purpose GeoAI. Section 6 presents empirical cases of bias detection in GeoAI audits, and Section 7 concludes with key findings and future research directions. 

\section{Background on bias}

\subsection{What is Geospatial AI?}

A frequently quoted but rarely verified claim says “about 80\% of data is geospatial.” The phrase originated in GIS trade literature and, although subsequent estimates differ (Barocas et al., 2023), the anecdote rightly signals geodata’s pervasiveness. With so much decision-making linked to location, even a modest systematic bias can have far-reaching consequences. Geospatial Artificial Intelligence (GeoAI) is a new interdisciplinary field where geospatial data and artificial intelligence are used to understand and gain knowledge about the human environment and the Earth. As \cite{gao2024introduction} describes, three major forces led to the fast development of this field: the development of AI tools, the availability of geographical information, and compute availability. Geospatial technology allows for making data-informed decisions in fields from urban planning and environmental management to public health and national security \autocite{world2020integrating, gevaert2024auditing}. 

Infrastructure for Spatial Information in Europe is a directive that drives the development of the EU-wide Spatial Data Infrastructure. Open Data Directive (2019/1024/EU) mandated that public and publicly-funded data with significant socioeconomic potential, like meteorological, statistical, and geospatial datasets, must be available for free. Figure 1 gives an overview of the data in the INSPIRE datasets, which illustrate the variety of geospatial data, ranging from administrative units, land cover, to species distribution. 
\begin{figure}
    \centering
    \includegraphics[width=0.8\linewidth]{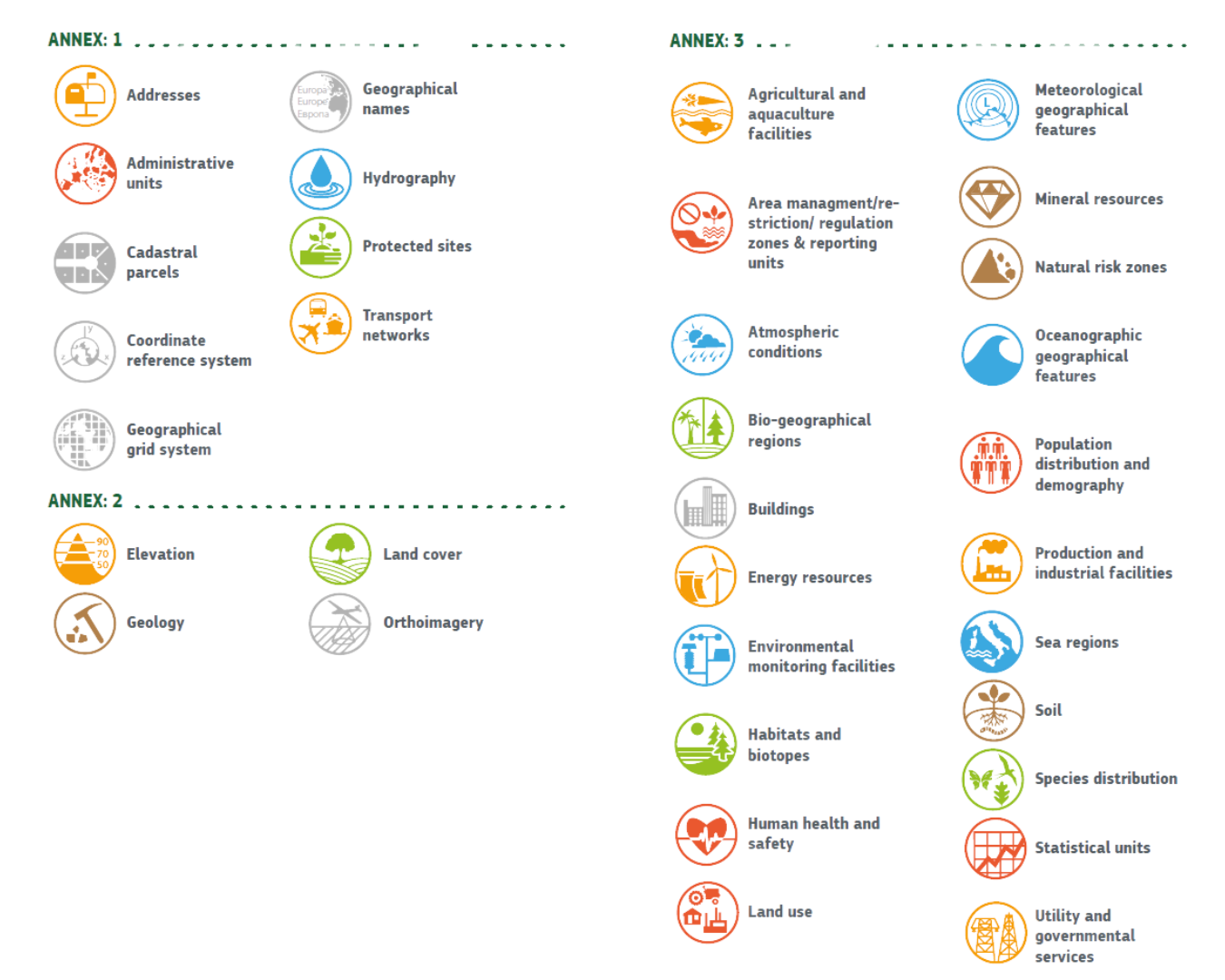}
    
     \caption{Data Themes in INSPIRE datasets.\\
    \textit{Note.} Reused from \textcite{Minghini2021}.
    Content is licensed under the {Creative Commons Attribution 4.0 International} (CC BY 4.0). 
    No content changes were made.
    }
    \label{fig:enter-label}
\end{figure}

The increasing availability of geospatial data allows to build AI systems used for optimization, learning, and decision making. \textcite{wang2024mapping} reviewed 415 peer-reviewed papers in quantitative human geography and organised them along two axes: data source (remote sensing imagery, street view photos, trajectories, social media, POI, census) and AI method (traditional ML, deep learning, NLP, graph models). The visual summary of their results is presented on Figure 2.
\begin{figure}
    \centering
    \includegraphics[width=0.8\linewidth]{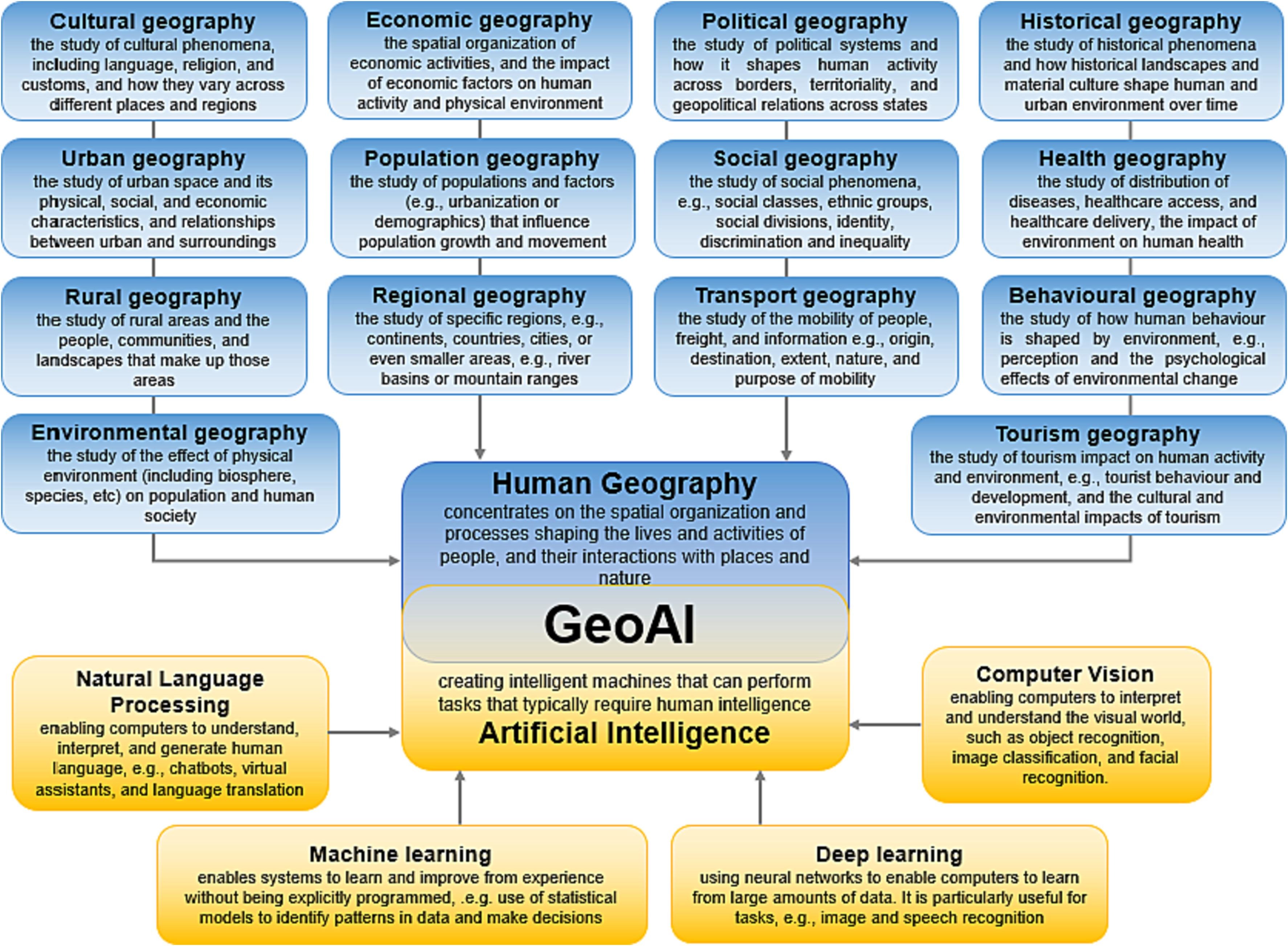}
    \caption{Types of geospatial data.\\
    \textit{Note.} Reprinted from \textcite{wang2024mapping}, 
    \textit{International Journal of Applied Earth Observation and Geoinformation}, \textit{120}, p.~2. 
    Content is licensed under the {Creative Commons Attribution 4.0 International} (CC BY 4.0). 
    No changes were made.
  }
    \label{fig:enter-label}
\end{figure}

GeoAI applications span public-value experiments as well as commercial analytics. Augmented Collective Intelligence is a framework by Dark Matter Labs that uses satellite imagery, open cadastral layers and citizen reports so local authorities can more accurately quantify the value of trees (including their role as stormwater and heat-mitigation infrastructure), not just street furniture \autocite{darkmatterlabsTreesInfrastructure}. On the private side, Retail Sonar geocodes loyalty card, census and footfall data to optimise growth strategies and predict store turnover. The two examples underscore GeoAI’s range, from climate adaptation budgeting to profit-seeking site selection, while foreshadowing a common risk: if input layers misrepresent certain districts or demographics, both public and commercial decisions can be skewed.

\subsection{How does bias arise in machine learning?}
\textcite{friedman1996bias} classified bias in computer systems into three categories: preexisting, technical, and emergent bias. Preexisting bias originates from societal structures, institutions, or individual prejudice and becomes embedded in the system through the training data or values of its designers. Technical bias arises from constraints in hardware, software, algorithms, or formalization choices that unintentionally lead to discriminatory outcomes. Lastly, emergent bias develops after deployment due to shifts in societal knowledge, user demographics, or cultural values that were not accounted for during design. As shown in Figure 3, \textcite{masinde2024auditing} uses this framework, as well as the work of Suresh and Guttag, to visualize how one can differentiate between types of bias, showing that "biases in early stages can have cascading effects" (p.8).
\begin{figure}
    \centering
    \includegraphics[width=1\linewidth]{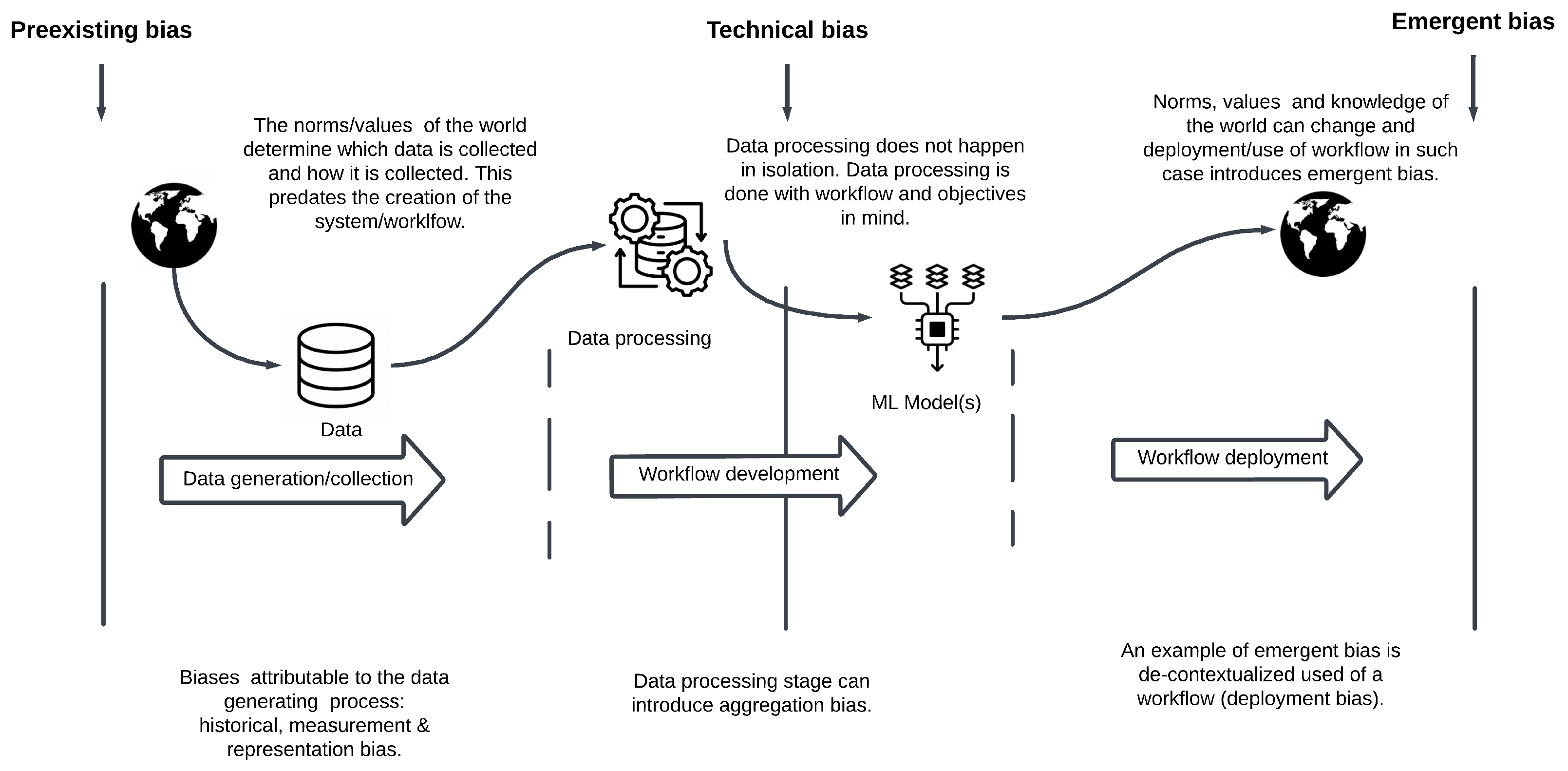}
     \caption{Nissenbaum-Friedman bias classification in AI workflows.\\
    \textit{Note.} Reprinted from \textcite{masinde2024auditing}, 
    \textit{ISPRS International Journal of Geo-Information}, \textit{120}(12), p.~8. 
    Content is licensed under the {Creative Commons Attribution 4.0 International} (CC BY 4.0). 
    No changes were made.
  }
    \label{fig:enter-label}
\end{figure}
These broad categories of bias can be further broken down, as demonstrated in \textcite{mehrabi2021survey} who built an extensive taxonomy of bias in machine learning. They showed that it is challenging to come up with a unified way of defining fairness, and that debiasing methods that emerge vary for specific applications and domains. Bias mitigation requires a contextual understanding of the data, the technical aspects of the model, its design decisions, and the way it was deployed. A simple example of (preexisting) bias is training Large Language Models (LLMs) on data biased towards minorities. \textcite{tamkin2023evaluating} evaluated Claude 2.0 for discrimination in various decision-making scenarios, detecting the model's bias towards marginalized groups using a discrimination score metric. They found that characteristics like coming from an ethnic minority, being female or non-binary increased the discrimination index, while age decreased it. Furthermore, the researchers tested methods to decrease this bias and found that prompt-based interventions improved performance. For example, stating that discrimination was illegal and asking the model to ignore demographic information decreased the discrimination score to zero or nearly zero. In the case of geospatial models, most mitigation techniques would need to rely on different methods, unless the model is an LLM. While the general principles of bias emergence apply broadly, GeoAI systems exhibit specific vulnerabilities due to the unique nature of spatial data. The next section focuses on bias mechanisms most relevant to GeoAI.

\subsection{Bias mechanisms specific to GeoAI}

The problem of bias in AI has been described in the literature for general applications and GeoAI specifically. Commonly discussed types of bias in geospatial AI include representation, aggregation, and algorithmic bias. Many more types of bias exist, as discussed in the previous section, but this paper focuses on these three due to their prevalence in the geospatial literature and relevance to the empirical cases presented in Section 6. 

Representation bias occurs when the population sample does not represent the real-world population well, and when the distribution of its subgroups is different. In particular, data collection methods can cause under-representation of certain groups. Collecting survey responses from social media users will misrepresent the total population, as users are predominantly young people \autocite{werner2021handbook}. This kind of bias can be detected by comparing group representation in the data to a real-world scenario, for example, by comparing the frequency of different age groups in the dataset and their actual world distribution. An example that highlights the potential issue in the European context is the availability of identifiable images from Open Images, 60\% of which are images from the U.S \autocite{shankar2017no}. The distribution of data sources is presented in Figure 4. Therefore, one can expect the performance of models like image classifiers, which are trained on such data, to underperform in the European context. This brings us to the problem of algorithmic bias, which means that an algorithm performs better in some types of areas. For example, if the model is mostly trained to detect buildings using data in cities, it can perform worse on rural data. Bias problems can further emerge as a result of design choices.
\begin{figure}
    \centering
    \includegraphics[width=1\linewidth]{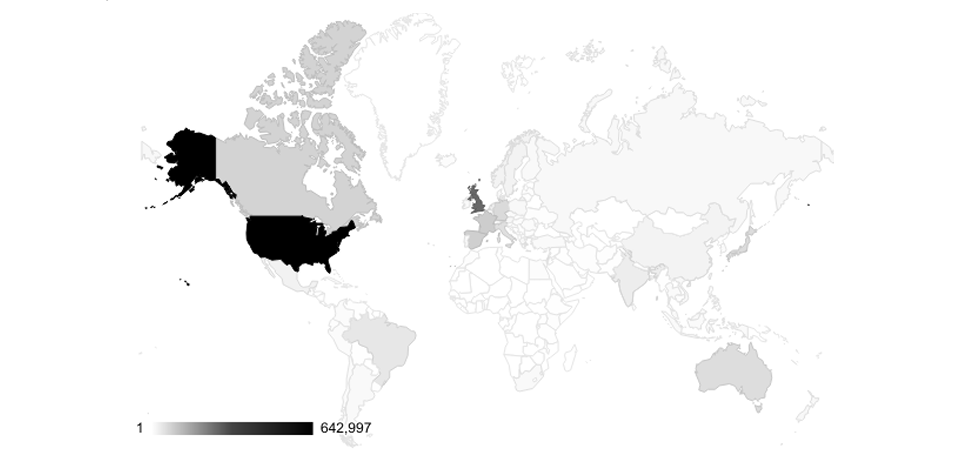}
     \caption{Distribution of identifiable images from the Open Images dataset.\\
    \textit{Note.} Reprinted from \textcite{shankar2017no}, 
    \textit{arXiv preprint arXiv:1711.08536}. 
    Used with permission of the author. No changes were made.
  }
    \label{fig:enter-label}
\end{figure}

Aggregation bias arises when we assume that trends observed in an entire population also hold for its distinct subgroups or individuals. By generalizing from the aggregate, we risk obscuring or even reversing important subgroup differences. An example of aggregation bias is demonstrated in the audit of \textcite{masinde2024auditing}, where Malawian houses were grouped into three categories. Yet, as authors discovered, within each category, the durability of individual house types varies significantly; this coarse classification disproportionately disadvantages the most vulnerable homeowners. A closely related issue is the modifiable areal unit problem (MAUP), in which the spatial patterns we detect depend entirely on the aggregation scheme we choose \autocite{MAUP}. An illustrative example is mapping disease density, which can hide areas with the highest disease spread by aggregating them with less-affected areas. This shows that in geospatial applications, researchers' aggregation choices can influence results just as much as the underlying data quality. Moreover, subsequent interpretation can matter even more: \textcite{lehner2008confirmation}, showed that personal judgments of analysts, and confirmation bias in particular, strongly shape how remote‐sensing data are understood.

\textcite{stevens2012mapping} identify data quality as a fundamental source of error in geospatial applications. The authors explain that whether data has high validity depends on whether it measures what it is supposed to measure, which in turn depends on precision, accuracy, and relevance. Accuracy is whether the measurement is close to its true value, and precision refers to the consistency of measurements, regardless of how close they are to the true value. Relevance reflects whether the data are appropriate for the task at hand. For example, location data correct to within one kilometer suffices for tracking wildlife but not for designing a missile defense system. \textcite{delmelle2022uncertainty} offers a concrete public health example of geospatial uncertainty. They show that measurement errors can be introduced with incomplete or wrong addresses of patients, geo-imputation (replacement of missing data), and data changes over time. These errors aggravate data quality and affect the analysis of disease prevalence, environmental exposures, and travel distances to care. Therefore, while geospatial data could seem to present the ground truth, the extent to which the data reflects the ground truth is often uncertain. These biases directly intersect with legal concerns under the EU AI Act, particularly in high-risk AI systems, where data representativeness and accuracy are explicit regulatory requirements (Article 10).

Applying machine learning and deep learning techniques to spatial data presents its own set of problems. Measurements, like temperature, taken in nearby locations tend to be similar, a phenomenon known as spatial autocorrelation. \textcite{ploton2020spatial} show that studies that ignore autocorrelation routinely overstate model performance in large-scale mapping applications. They note that ignoring spatial autocorrelation in evaluations for model performance is a common practice, suggesting that this problem remains unchecked. Besides being spatially autocorrelation, the real-world data is also imbalanced, due to the lack of data for certain regions and rare phenomena, like oil spills in water images \autocite{koldasbayeva2024challenges}. \textcite{shaban2021deep} build a detection algorithm trained on severely imbalanced data, where oil spill instances represent only 1.2\% of pixels. Non-uniform data distribution poses a challenge to traditional training models, which led to the development of numerous techniques addressing this problem \autocite{krawczyk2016learning}. Scale dependency adds another layer of complexity: ecological patterns and processes can vanish or appear differently when viewed at different spatial or temporal scales \autocite{hewitt2010scale}. The variety and prevalence of bias in machine learning and the presence of specific geospatial problems highlight the importance of developing context-specific mitigation methods. Existing applications in public safety, disaster management, and environmental analysis demonstrate how geospatial AI directly intersects with Article 6 (high-risk classification) and others in the EU AI Act, where bias can compromise safety, fairness, and fundamental rights. 

\section{Evidence of the European data bias research gap}

Masinde et al. observe that “systematic auditing for bias is still nascent in GIScience and disaster-risk management” (2024, p. 2). Our own literature search supports that point: we located only a small set of published European spatial data audits, such as layer-specific tests on vegetation \autocite{vegetation} or impervious-surface indices \autocite{strand2022impervioussness}, and virtually none examine the use of GeoAI. Yet, as Section 2 showed, bias is well documented across spatial domains, and flagship EU datasets are not immune. For instance, provenance statistics indicate that roughly 92\% of all INSPIRE datasets originate from German sources, which is presented in Appendix A. This observation does not prove bias, but it does suggest that German data is disproportionately easier to obtain, raising the question of how a model trained on such material will perform elsewhere in Europe. Masinde et al. and a recent European Parliamentary Research Service (ERPS) report—Auditing the Quality of Datasets Used in Algorithmic Decision-Making Systems \autocite{de2022auditing}—argue that routine, transparent audits are essential. The EPRS study explicitly recommends dataset audits as a policy lever to enforce fairness requirements, noting that without them, “representativeness problems remain invisible to regulators and users” (p. 19).

The current lack of public audits of the European datasets and GeoAI, despite the prevalence of bias in geospatial applications, is the primary motivation for this paper to examine how the EU AI Act addresses this issue. The EU AI Act addressed bias with obligations throughout Articles 9–15: developers of high-risk systems must analyze foreseeable risks, document data representativeness, and monitor post-deployment drift. Section 4 unpacks those provisions and links them to the problem of bias in GeoAI. Section 5 then turns to practical bias mitigation, presenting three GeoAI audit case studies and outlining approaches that can make such evaluations part of a normal development workflow. Before examining practical mitigation and audit case studies, it is crucial to understand the specific legal obligations imposed by the EU AI Act, which aims to address these identified gaps and risks. 

\section{Legal framework: EU AI Act and GeoAI}

The EU AI Act entered into force in August 2024, and its obligations for high-risk systems apply 36 months after entry into force (August 2027). The Act defined a risk-based approach to building safe AI and is the first piece of horizontal legislation by a major policymaker that comprehensively regulates the development of AI products \autocite{europaActFirst}. It significantly impacts the development of AI systems, as depending on the level of risk posed by the system, it places legal requirements on the developers of AI systems. The Act serves the goal of mitigating risks posed by such systems to protect fundamental human rights. This section describes the trustworthy AI principles that the Act is based on, the risk-based approach and provisions related to the problem of bias.

\subsection{Trustworthy AI Principles}
The 2019 Ethical Guidelines for Trustworthy AI, published by the European Commission's High-Level Expert Group on Artificial Intelligence (AI HLEG), are referenced in the EU AI Act because they provide the foundational ethical principles that guide the legislation's risk-based framework. These requirements include: human agency and oversight; technical robustness; privacy \& data governance; transparency; diversity, non-discrimination \& fairness; societal \& environmental well-being, and accountability. According to Recital 27, these principles are all equally important and should be evaluated throughout the model's lifecycle by stakeholders \autocite{EUAIActFinal2024}. It should be noted here that Recitals, such as Recital 27, serve to interpret and provide reasoning behind the Articles. These seven principles inspired the approach of the AI Act, and while their wide voluntary adoption is encouraged, they were also translated into binding obligations for high-risk AI under Articles 8 to 15. Appendix B presents an overview of this framework.

\subsection{Classification of AI systems using the Risk-Based Approach}

Building on these principles, the Act’s risk-based approach sets different requirements for AI systems depending on their potential to cause harm. AI systems are segmented into four tiers of risk: unacceptable risk, high-risk, low-risk, and minimal risk. Unacceptable risk category includes systems with a level of risk prohibited in the EU, including subliminal manipulation and social scoring. High-risk systems include systems that can significantly impact human well-being or fundamental rights, and are thus regulated. Limited-risk systems do not fall under high-risk or prohibited categories but potentially pose smaller risks (like chatbots that could manipulate decision-makers), and therefore face lighter obligations. Systems posing minimal risks, like basic recommendation systems, are viewed as posing no risk to safety and can be deployed without additional requirements.

Article 6 specifies that AI products are classified as high-risk when they fall under existing EU product safety rules with harmonized risk assessments or serve applications listed in Annex III \autocite{EUAIActFinal2024}. The areas listed in the Annex include biometrics, critical infrastructure, education and vocational training, employment, access to private and public services, law enforcement, migration, and administration of justice and democratic processes. To give an example, Recital 61 explains why certain AI products intended for the administration of justice and democratic processes are classified as high-risk. Such systems can impact "democracy, rule of law, individual freedoms", contain biases, and produce errors. Therefore, both systems assisting decisions and those serving as alternative dispute resolution bodies should be treated as high-risk. They highlight the importance of human decision-making and, at the same time, clarify that systems supporting purely administrative tasks are considered low-risk. To better explain how geospatial data could be used in systems that are considered high-risk, Table 1 lists concrete examples of such systems. This is critical for AI developers to understand, as they need to adhere to requirements specified in the EU AI Act.

\begin{table}[H]
    \centering
    \begin{tabular}{p{4cm}p{6cm}p{5cm}}
        \toprule
        \textbf{EU AI Act High-Risk Classification} & \textbf{Geospatial Example} & \textbf{Why It Matters} \\
        \midrule
        Law Enforcement & Predictive policing using spatio-termporal analysis with learning models, developed for by German police departments \autocite{egbert2024algorithmic}& Bias in spatial data may amplify existing societal inequities by promoting over- or under-policing of specific neighborhoods, as shown by \autocite{richardson2019dirty}.\\
        \midrule
        Critical Infrastructure & Disaster management using GeoAI for applications like building detection, climate solutions and flood vulnerability maps, like  FloodAI \autocite{engagementhqFloodAI} & Misrepresentation of certain regions (e.g., rural or poor areas) can deny resources or delay aid, violating fairness and safety provisions.\\
        \midrule
        Access to Public Services & Automated allocation of public utilities and resources, e.g., TreesAI for optimizing urban tree planting benefits. \autocite{darkmatterlabsTreesInfrastructure} & Biased geospatial datasets (e.g., incomplete building footprints) risk excluding marginalized communities, breaching non-discrimination requirements.\\
        \midrule
        Administration of~Justice& Forensic geospatial analysis and crime mapping using GIS tools (that often use machine learning techniques) to analyze spatial evidence for legal proceedings \autocite{arcgisOverviewCrime}.& Geospatial bias in data could skew legal boundaries or resource distribution, potentially infringing on fundamental rights and fair trial standards. \\
        \midrule
        Education and Vocational Training & School geocoding, travel-distance mapping, and kernel density estimation to analyze disparities in access to education across urban districts \autocite{cobb2020geoeducation}.& Overlooking certain geographic areas (e.g., lower-income or remote regions) could reinforce educational inequities, conflicting with diversity/fairness principles. \\
        \midrule
        Migration and Border Control & Geospatial risk assessment and border surveillance, like Frontex Surveillance Systems developing AI for marine awareness \autocite{europaPROMENADEArtificial}.& Incorrect or biased location data may mislabe zones, affecting fundamental rights and the safety.\\
        \midrule
        Democratic Processes &  Existing AI methods for creating electoral districts  and analyzing potential gerrymandering \autocite{cho2023ai}.& Algorithmic bias in redistricting tools could manipulate district boundaries, undermining democratic fairness.\\
        \bottomrule
    \end{tabular}
    \caption{Geospatial Bias Risks in High-Risk AI Applications Under the EU AI Act}
    \label{tab:geospatial_bias}
\end{table}
Note. GDPR (Article 4(14)) defines biometric data as "personal data resulting from specific technical processing relating to the physical, physiological or behavioral characteristics of a natural person, which allow or confirm the unique identification of that natural person, such as facial images or dactyloscopic data." Therefore, even though one can imagine biometric data with a location component and even though privacy breaches are an issue in geospatial data \autocite{EthicsPrivacy}, this problem is outside the scope of this paper.

While Table 1 below highlights a few concrete GeoAI examples that fall into high-risk categories, it represents only a narrow slice of the broader landscape. As Copernicus data is used to train AI \autocite{europaSupportingEurope}, and as established geospatial platforms like ArcGIS are constantly improving their GeoAI applications, \autocite{arcgisTrustedArcGISx2014Trusted}, one might expect GeoAI to become increasingly important in critical areas. Agencies, including Europol's Innovation Lab, announced plans to expand AI use in border control and law enforcement \autocite{europol2023report}, highlighting the importance of bias mitigation in their report. This table gives existing public examples of GeoAI in critical areas, and as this technology develops, it will likely have increasingly wider applications for institutions and states.

\subsection{Requirements for high-risk systems}

As previously described, providers of systems classified as high risk face specific requirements concerning different parts of the life cycle and parts of their system. Specifically, Article 9 requires providers, meaning entities that place the systems on the EU market, of high-risk systems to create and maintain a risk-management system. They must analyze foreseeable risks of their system, evaluate them, and adopt necessary measures to mitigate these risks.

Article 10 sets binding legal requirements for data governance and data quality. It details the requirements related to data governance, focusing on data quality requirements during training, validation, and testing. The data governance and management practices are relevant for, among others, the process of collecting data, data preparation, including "annotation, labeling, cleaning, updating, enrichment and aggregation," and data quality assessments. The article specifically mentions the process of examining data for possible biases that could impact fundamental rights and designing measures to prevent and mitigate them. Moreover, the datasets need to be "relevant, sufficiently representative", as free of errors as possible, and complete for the intended purpose (EU AI Act, Art. 10(3)). Recital 67 clarifies why data representativeness is central to non-discrimination, stating that AI systems should not disproportionally harm and exclude specific groups. To avoid biases and ensure the suitability of the model for its purpose, datasets should be designed to reflect the specific geographical, contextual, behavioral, or functional settings in which the AI system will be used. This means that AI models should be trained on data that accurately represents the real-world conditions in which they will operate, ensuring relevance and reducing biases.

Next, Article 11 requires providers to compile technical documentation to show their compliance with the Act. The minimal required documentation is listed in Annex IV, and includes information on the data used, the development of the AI system, as well as its potential discriminatory impacts and foreseeable sources of risks to health, safety, and fundamental rights. Bias mitigation is not a simple checkmark mentioned in Article 10, but it is woven throughout the obligations on providers of high-risk AI, like in Articles 13 and 14. 

Article 13 focuses on transparency obligations, requiring providers to supply clear instructions for use and other relevant information, such as the intended purpose of the AI system, its performance and accuracy, and any operational constraints. Although Article 13 does not explicitly reference bias, its transparency requirements can help users identify potential limitations or unintended effects. To give an example, by stating where or under what conditions the system has been tested, providers can alert users if the model underperforms for certain rural or underrepresented regions, making it easier to spot any overlooked patterns of discrimination. Article 14 requires that high-risk AI systems be designed and developed in a way that allows effective oversight of human operators. Human oversight allows for monitoring the capacities and limitations of high-risk systems, and notice over-reliance on decisions made by the high-risk system. The goal is to be able to correctly interpret and decide whether to use or override the output of the system. 

Article 15 mandates that high-risk AI systems must achieve an appropriate level of accuracy, robustness, and cybersecurity. It requires the developers to maximally reduce the risk of biased outputs that could lead to negative feedback loops. The article underlines the importance of detecting inherent bias in data, as it could lead to negative feedback loops. This phenomenon is relevant in the case of geospatial AI, as groups that are underrepresented in the dataset face harmful consequences. An example is building detection AI, that, as a result of being trained on mostly urban datasets, incorrectly detects buildings in rural regions \autocite{gevaert2024auditing}. Consequently, rural households can become invisible to decision makers, further worsening their situation.

Recital 75 describes technical robustness as a "key requirement for high-risk AI systems". It highlights that providers are responsible for making their systems resilient to errors and unexpected situations like data inconsistencies. For example, systems could include fail-safe mechanisms that detect anomalies and safely interrupt operations when necessary. Generally, safeguards need to prevent and minimize harmful behavior that could negatively impact fundamental rights by producing wrong or biased outputs, and erroneous decisions. Recital 70 states that AI providers may process special categories of personal data (such as race, ethnicity or health data) in exceptional cases to protect from discrimination. Under appropriate safeguards and compliance with other EU regulations, including GDPR, they may use this data to detect and reduce bias in AI systems. 

Table 2 summarizes how the Act’s life-cycle duties (Arts. 9–15) relate to the five bias mechanisms introduced in Section 2.3. For each bias, the gatekeeper article is the first clause that obliges the provider to prevent or document the error source; secondary duties such as transparency (Art. 13) or human oversight (Art. 14) follow once the gatekeeper task is complete. This mapping is intended as a practical summary, not a legal opinion, and should be refined for any specific deployment. 

\begin{table}[H]
\centering

\begin{tabular}{>{\raggedright\arraybackslash}p{4cm} >{\raggedright\arraybackslash}p{6cm}>{\raggedright\arraybackslash}p{5cm}}
Bias mechanism& Gatekeeper article \& clause(s)& Why this is the first-line obligation \\
\toprule
\textbf{Representation bias} (data collection \& curation stage) & \textbf{Art. 10 § 2 (a-d)} – data must be “relevant, representative, free of errors, complete”;Recital 67 on geographical representativeness & Article 10 is the only provision that explicitly demands representative training data and mandates checks for biases impacting fundamental rights.\\
\midrule
\textbf{Aggregation bias} (feature engineering, spatial aggregation, evaluation) & \textbf{Art. 9 § 2 (a-c)} – risk-management system must identify, analyse, evaluate foreseeable risks throughout the life-cycle & Unexamined aggregation is a possible source of discriminatory error; risk analysis must consider alternative spatial scales and subgroup effects.\\
\midrule
\textbf{Algorithmic bias} (training \& tuning; post-deployment)& \textbf{Art. 15 § 1-2} – high-risk AI must achieve appropriate accuracy and robustness; Art. 15 § 4 requires post-market monitoring& Spatially uneven accuracy constitutes a robustness failure; negative feedback loops arise when poor performance systematically harms under-represented areas. \\
\midrule
 \textbf{Measurement / labelling bias} (sensor error)& \textbf{Art. 15 (accuracy \& robustness)} (and Art. 10 “free of errors”)&Systematic sensor or annotation error compromises accuracy before any modelling.\\
 \bottomrule

\end{tabular}

\end{table}

Therefore, the aim of bias reduction is to reduce harms to health, safety and fundamental rights that could stem from discriminatory outputs of high-risk systems. In the EU AI Act, bias reduction is a key step to design safe AI systems and avoiding discriminatory bias. Multiple articles formally require the evaluation of discriminatory impacts, bias detection, and mitigation from the providers of high-risk AI systems.

\section{Future outlook: General Purpose GeoAI}

The EU AI Act also concerns the developers of general-purpose AI (GPAI) models, which it warns may pose systemic risks across different domains like democracy, security, and safety \autocite{EUAIActFinal2024}. These risks arise from misuse, fairness issues, cybersecurity threats, physical system control, and the spread of misinformation and require global attention and regulation to prevent large-scale harm. The documentation requirements in Annex XI specify that developers of such models similarly need to report methods that were used to identify biases. According to the classification described in Article 51, AI systems are considered GPAI when they have broad, multi-domain applications, serve as a foundational model for other systems, have potential systemic impacts, and operate with high autonomy. 

In the geospatial arena, major platforms like ArcGIS, begun increasingly integrating machine-learning-based geospatial analytics, offering pretrained models for tasks like object detection and feature extraction in satellite imagery. However, these applications remain narrowly targeted rather than true AI foundation models. Researchers proposed multi-scale or multi-level language models that utilize spatial representations to potentially serve as “backbones” for more flexible geospatial tasks, yet these efforts are still nascent compared to the breadth of general-purpose LLMs used for general text analysis \autocite{kulkarni2021multi}. Consequently, while specialized 'GeoAI' frameworks continue to mature and show promise in fields such as geoparsing, risk assessment and location-based analytics, a widely adopted general-purpose GeoAI that is on par with foundation LLMs in other domains has not yet fully materialized. In their survey, \textcite{chen2023foundation} mention the trend towards building foundation models for earth observations. Several projects are underway to create versatile, large-scale models that, once trained, can be fine-tuned for a variety of applications.

\section{How is bias detected in audits? Empirical cases and general rules}

Generally, the first step to identify bias in AI systems is to identify its potential sources in training data, algorithms, and user interactions, keeping in mind its common sources. \textcite{mehrabi2021survey} explain that the various methods for bias detection can be divided into techniques that transform data to remove underlying discrimination, techniques that modify learning algorithms to remove discrimination during the learning process, and post-processing methods that are performed after training. The last one is especially useful when the model is treated as a black-box, like in the  case of LLMs. Generally, while there are numerous methods, they mostly rely on evaluating the model's predictions across different subgroups and comparing outcomes to identify any disparities. The chosen method can depend on the preferred definition of fairness, the context, and can also be chosen by comparing the effectiveness of detection tools, like  AI Fairness 360 or Aequitas \autocite{mehrabi2021survey}.

Recent empirical work shows why such audits matter for spatial data. \textcite{gevaert2024auditing} conduct an audit of open building datasets to assess bias in the disaster management context. They use datasets of buildings in the Philippines and Tanzania, revealing bias in the accuracy of data based on sensitive attributes (such as the level of poverty in the area). The authors focus on auditing open building delineation datasets (Google Open Building, Microsoft's Building Footprints) that were created using deep-learning algorithms and where bias could have been introduced at different stages. The authors create the reference data manually and include a variety of areas, like urban, rural, poorer, and more populated areas. They compare the global datasets against the reference data and find large amounts of false negatives (missed buildings) and false positives (incorrect buildings). 

\textcite{masinde2024auditing} audit an AI-based flood vulnerability system in Malawi, identifying aggregation and representation bias. They found that data limitations and workflow design introduced bias at multiple stages of the GeoAI development, showing that it is crucial to assess each step individually. The study reveals that thatched roof buildings are underrepresented in the training data, leading to lower detection rates compared to metal roofs. Similarly, aggregating building types by classifying walls as either brick or concrete results in 26\% of houses being misclassified due to alternative construction materials. Bias persists in later stages when buildings are grouped into broad categories, further obscuring underrepresented structures. These errors distort flood risk assessments, potentially denying aid to disadvantaged housing types. Given these flaws, the authors recommend against implementing the system, as it fails to accurately represent vulnerable groups. They also highlight the risk of unintentional discrimination, as certain building types (often tied to ethnicity or tradition) are systematically overlooked.

\textcite{liu2022geoparsing} examine geoparsing systems, which turn unstructured data (like news articles or reports) into geospatial data. These systems help create heatmaps, travel routes, and incident maps. The authors note that, thanks to the use of deep learning techniques, there have been significant improvements in geoparsing performance. Their goal is to check whether this high performance is uniform over regions or if there are disparities by evaluating four state-of-the-art geoparsing models for toponym recognition and resolution. They test on 2122 geo-annotated tweets and three resolution benchmarks (LGL, GeoVirus, WikToR). They find that over 50 \% of place mentions are missed and that median resolution errors exceed 1 000 km. Spatial clustering shows “hot” areas only in the US, UK, and Malaysia, while “cold” regions cover South America, India, and the Middle East, showing inconsistent accuracy. Therefore, they conclude that previous performance evaluations have been overly optimistic, since they focus on overall performance rather than spatial variation. This may become a bigger problem as the growth of LLM accelerates, and so does the ease with which one can turn unstructured data into structured data with LLMs.

These empirical findings underscore the importance of data auditing and context-sensitive performance metrics. Although the empirical cases demonstrate successful methods for detecting bias in GeoAI, such public audits remain scarce. Under the EU AI Act, addressing these biases is no longer a mere best practice, but instead, becomes a regulatory requirement for high-risk systems. This highlights the need for developing tools and audit methods that can be used to ensure ethical and compliant deployment of GeoAI in the EU.

Market-based solutions may further help with this problem in the future. While there are existing standards and tools like Aequitas and IBM 360 \autocite{bellamy2018ai, saleiro2018aequitas}, more such tools will likely appear, as governments are interested in growing their AI Assurance markets. A 2024 UK Department for Science, Innovation \& Technology report estimates that the current AI-assurance market captures only a fraction of potential demand and highlights “testing, audit and certification services” as priority growth areas \autocite{undefinedAssuringResponsible}. Extending frameworks with spatial-specific metrics and integrating them into AI workflows remains an open research and commercial opportunity.

\section{Conclusion}

This paper offered the first integrated review of how the EU AI Act reshapes obligations to detect and remedy bias in Geospatial AI. GeoAI is increasingly central to various applications, from urban planning and resource allocation to disaster response and law enforcement. We discussed evidence of bias in GeoAI systems and mapped specific bias mechanisms onto the Act's risk-management and governance obligations. 

Our analysis makes four key contributions: (1) a concise overview of representativeness, aggregation and algorithmic bias, as well, as GeoAI-specific manifestations (spatial autocorrelation, MAUP); (2) a novel classification table demonstrating that currently deployed GeoAI services meet the EU AI Act’s definition of high‑risk systems; (3) a table linking each bias mechanism to relevant governance provisions in Articles 9-15; (4) an outline of how bias can be detected and mitigated with examples of GeoAI audits. These outputs address the research goals from Section 1 and confirm that technically detectable biases already affect high-risk GeoAI systems.

However, because our conclusions rely on general evidence of bias in GeoAI and on only a few European studies rather than direct proof in high-risk GeoAI, they remain provisional. Although Chapter 3 identified some European geodata audits and we performed a basic check of INSPIRE provenance, routine, targeted audits are necessary to detect and mitigate bias in European GeoAI. This view is shared by the European Parliamentary Research Service's and Europol's reports, which both explicitly mention the need for auditing and bias detection \autocite{de2022auditing, europol2023report}. It then follows that GeoAI audits are scarce despite their recognized value. As \textcite{masinde2022algorithmic} notes, “systematic auditing for bias is still nascent in GIScience and disaster-risk management” (2024, p. 2), reflecting that both the field and its auditing methods are still evolving. Likewise, \textcite{ploton2020spatial} observes that ignoring spatial autocorrelation tests can overstate model performance, and \textcite{liu2022geoparsing} shows that existing geoparsing performance metrics become overoptimistic when spatial variation is overlooked. These studies suggest that GeoAI developers might be incentivized to overstate their models' performance by avoiding additional, time-consuming evaluations. Ultimately, these findings imply that progress in bias detection will depend less on new theoretical advances than on the steady normalization of transparent, context-specific audits. The EU AI Act's full entry into force in 2027 is a step in this direction, as it creates legal requirements that make bias mitigation obligatory.

Future research should concentrate on several key areas to further ensure the ethical and responsible development of GeoAI. A significant gap exists in the auditing of EU geospatial datasets and GeoAI systems for bias, and this should be a priority for future work in the European context. There is also a need for the development of context-specific auditing tools and methods to effectively assess and mitigate bias in GeoAI applications. Finally, as general-purpose GeoAI models emerge, their potential systemic risks and implications for bias require thorough investigation. Overcoming these challenges is essential to building GeoAI systems that are powerful, fair, accountable, compliant with the EU AI Act, and beneficial for humanity’s future.

\break

\printbibliography
\break
\appendix

\section{Appendix A}
\begin{figure}[H]
    \centering
    \includegraphics[width=0.5\linewidth]{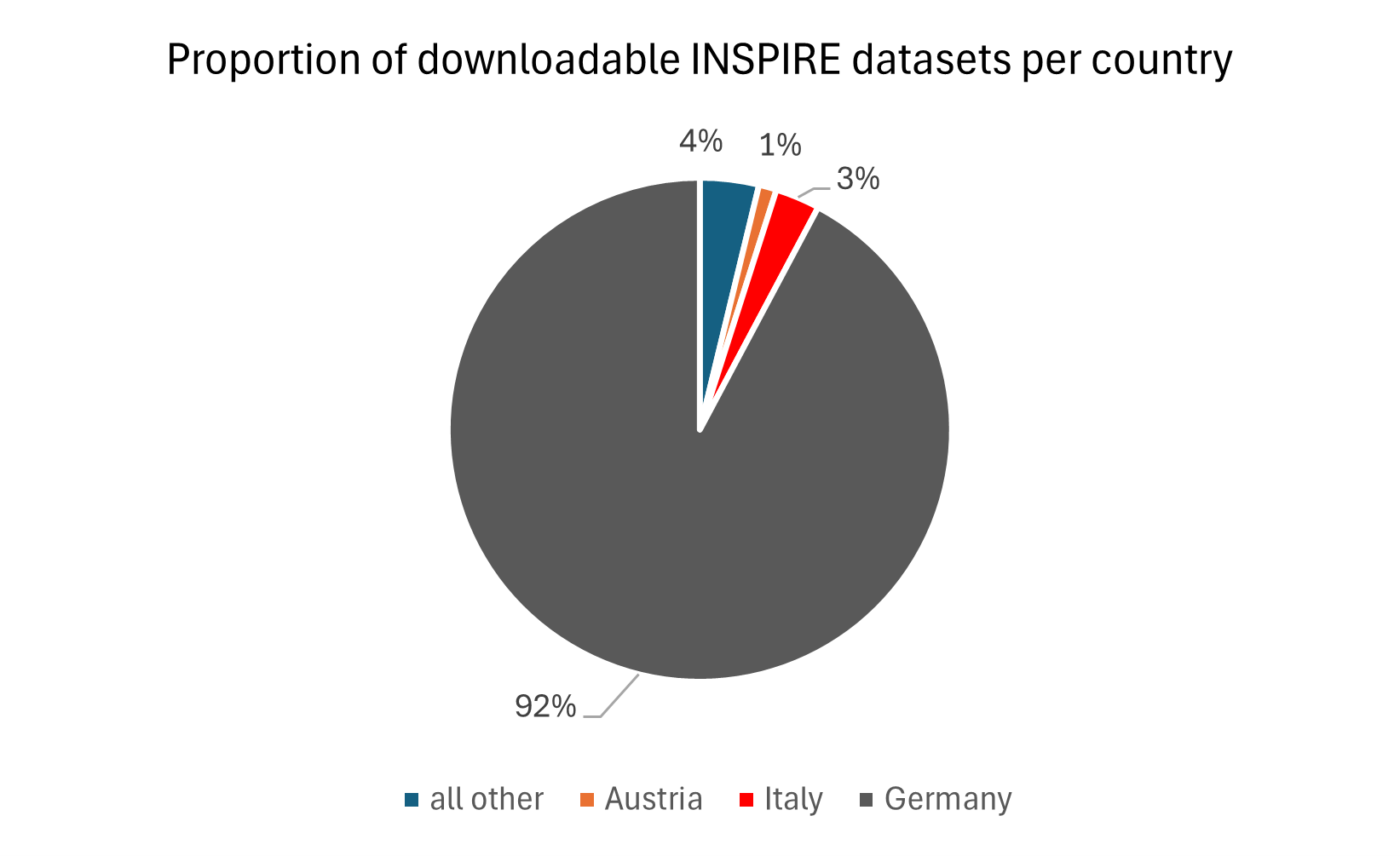}
    \caption{Proportion of INSPIRE downloadable datasets. Source: \autocite{europaINSPIREGeoportal}}
    \label{fig:enter-label}
\end{figure}
\textit{Note}. Percentages were calculated in Excel by dividing the count of German-origin datasets by the total number of INSPIRE downloadable datasets. 

\break

\section{Appendix B}
\begin{figure}[H]
    \centering
    \includegraphics[width=1\linewidth]{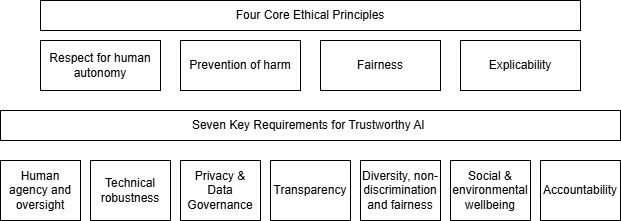}
    \caption{Ethical principles and requirements in 2019 Ethical Guidelines for Trustworthy AI.\\
    \textit{Note.} Based on \textcite{AIHLEG2019}. The EU AI Act refers to the seven 'requirements' as 'principles', and does not reference the four principles in the first row.
  }
    \label{fig:enter-label}
\end{figure}

\end{document}